\documentclass[useAMS,usenatbib]{mn2e}
\usepackage{epsfig}
\usepackage{amsmath}
\usepackage[firstpage]{draftwatermark}

\newcommand{\xmm}{\textit{XMM-Newton}\ }
\newcommand{\chandra}{\textit{Chandra}\ }

\newcommand{\etal}{et al.\ }

\usepackage{epstopdf}

\title[Optically selected fossil groups]{Optically selected fossil groups; X-ray observations and galaxy properties}
\author[Khosroshahi et al.]{Habib G. Khosroshahi$^{1}$\thanks{E-mail: habib@ipm.ir},
Ghassem Gozaliasl$^{2,3}$, Jesper Rasmussen$^{4}$, \newauthor  Alireza Molaeinezhad$^{1}$, Trevor Ponman$^{5}$, Ali A. Dariush$^{6}$, Alastair J.R. Sanderson$^{5}$\\
$^{1}$School of Astronomy, Institute for Research in Fundamental Sciences (IPM), P. O. Box 19395-5746, Tehran, Iran \\
$^{2}$School of Physics, Tabriz University, Tabriz, Iran \\
$^3$Department of Physics, University of Helsinki, P. O. Box 64, FI-00014 Helsinki, Finland\\
$^{4}$Dark Cosmology Centre, Niels Bohr Institute, University of Copenhagen, Juliane Maries Vej 30, DK-2100 Copenhagen, Denmark\\
$^{5}$School of Physics and Astronomy, University of Birmingham, Birmingham B15~2TT, UK\\
$^{6}$Institute of Astronomy, University of Cambridge, Madingley Road, Cambridge CB3 0HA, UK}

\begin{document}
%%%%%%%%%%%%%%%%%%%%%%%%%%%%%%%%%%%%%%%%%%%%%%%%%%%%%%%%%%%%%%%%%%%%%%%%%%%%%%%%%%%%%%%%%%%%

\maketitle

\label{firstpage}
%*******************************************************************************************
\begin{abstract}

We report on the X-ray and optical observations of galaxy groups selected from the 2dfGRS group catalog, to explore the possibility that galaxy groups hosting a giant elliptical galaxy and a large optical luminosity gap present between the two brightest group galaxies, can be associated with an extended X-ray emission, similar to that observed in fossil galaxy groups. The X-ray observations of 4 galaxy groups were carried out with \chandra telescope with 10-20 $ksec$ exposure time. Combining the X-ray and the optical observations we find evidences for the presence of a diffuse extended X-ray emission beyond the optical size of the brightest group galaxy. Taking both the X-ray and the optical criteria, one of the groups is identified as a fossil group and one is ruled out because of the contamination in the earlier optical selection. For the two remaining systems, the X-ay luminosity threshold is close to the convention know for fossil groups. In all cases the X-ray luminosity is below the expected value from the X-ray selected fossils for a given optical luminosity of the group. A rough estimation for the comoving number density of fossil groups is obtained to be 4 to 8$\times10^{-6}$ Mpc$^{-3}$, in broad agreement with the estimations from observations of X-ray selected fossils and predictions of cosmological simulations. 

\end{abstract}
%********************************************************************************************
%********************************************************************************************
\section{Introduction}

Fossil groups \citep{ponman94} are galaxy groups dominated by a single giant
elliptical galaxy, some as luminous as brightest cluster
galaxies (BCG). Formally, fossil groups are identified as having 
X-ray luminosities comparable to X-ray bright groups ($L_X{\ge}10^{42}{\rm
erg\,s}^{-1}$) and a difference of 2 magnitude or more between the
first and second ranked galaxies within half the group Virial
radius \citep{jones03}. The physically origin of fossil galaxy groups is argued to be in galaxy mergers. 
The X-ray observations of galaxy groups show that some of these groups have substantial 
amount of  group scale X-ray emission which itself indicates deep gravitational potential of dark matter. 
This indicates that the hot gas is retained during the process of galaxy mergers. 

There are now tens of fossil groups identified \citep{kjp04,sun04,ulmer05,kmpj06,santos07} observationally 
mostly meeting the criteria introduced by \citet{jones03}. Due to their observationally distinct 
properties relative to the general population of galaxy 
groups, they are believed to be suitable systems to study the formation and evolution of giant elliptical 
galaxies \citep{kpj06,smith10} and galaxy groups and clusters  \citep{kpj07,dariush07}. Our  earlier 
\chandra study of a small sample of fossil groups \citep{kpj07} shows
several interesting features. The X-ray morphology of fossils is
undisturbed indicating the absence of recent major merger. The
dominant galaxy lies at the centre of group-scale X-ray emission, 
which traces the dark matter potential well.  In addition, fossils are 
over-luminous in X-rays, for their optical luminosity, but fall comfortably on the standard
$L_X-T_X$ relation \citet{kpj07} suggesting that their hot gas temperature
may also be elevated. The dark matter distribution in fossil groups is
more centrally concentrated than in other groups and clusters, indicating early formation epoch for fossils.  
The differences between fossils and non-fossil groups are not limited
to their hot gas and dark matter. The dominant central
galaxies in fossils have disky or pure elliptical isophotes, while the
majority of giant elliptical galaxies, and BCGs in particular, have
boxy isophotes \citep{kpj06, smith10}.

The observed properties of fossil groups point to their early formation and lack of later
disturbance. Thus they must be amongst the first
systems to have been virialized, which makes them very important probes
of the early Universe. In many respects they should be similar to
the cores of clusters, but with the distinction that fossil groups
have not been subject to the subsequent subcluster merging which
leads to the growth of galaxy clusters. These distinctive characteristics makes fossil groups especially valuable
as a check on the adequacy of cosmological simulations. We have shown, using
the Millennium simulations \citep{springel05}, that galaxy groups with large luminosity gaps are formed
relatively earlier than groups with least luminosity gap between the two brightest galaxies \citep{dariush07,dariush10}. Also fossils are seen as suitable environments to study IGM heating and cooling in the absence of group scale mergers \citep{miraghaee13}. The luminosity gap as an observable quantity for a galaxy group is also shown to be a strong probe of galaxy formation and evolution models \citep{b17, gozali13}. 

The primary aim of this study is to test the hypothesis that most
of the galaxy merging which builds up giant elliptical galaxies actually takes
place in collapsed groups. If this hypothesis is true, one could also  
expect that a great majority of purely optically selected 
fossil groups, e.g. groups with large luminosity gap between the two brightest members, 
to show group-scale X-ray emission. If this turns out to be true, it would result in a major 
advance in understanding the formation of luminous elliptical galaxies. In addition it would 
significantly raise the number of known fossil groups, and establish an
economical and efficient method for finding more such systems. 

To achieve this, we carried out \chandra observations aimed at detecting X-ray emission from optically selected groups. \cite{jesper06} adapted a similar approach using the \xmm and found low X-ray luminosity ($L_X\approx10^{40} {\rm erg\,s}^{-1}$) in several groups which were selected at random from the 2dFGRS to have group velocity dispersion, $\sigma_v\approx 300$ km~s$^{-1}$. They argued that their groups may have not been fully collapsed, as an explanation of their findings. However, their sample had no constraint on the brightest group galaxy luminosity/morphology or on the magnitude gap between the two most luminous  member galaxies, which feature in our selection of galaxy groups. 

In this contribution, we report on the X-ray and the optical observations of the sample. Section 2 describes our sample selection and observation strategy, the X-ray and the optical observations. Section 3 is dedicated to the X-ray analysis followed by optical properties of the groups and the galaxy luminosity function in section 4. In section 5, a discuss of the results and our conclusions are given.  $H_0 = 70$ km s$^{-1}$ Mpc$^{-1}$ and  $\Omega_m = 0.3$ are assumed throughout this paper. 

\section{Sample selection and observations}

To address the question of whether optically selected groups dominated by luminous elliptical galaxies (e.g. groups with a large luminosity gap between the two brightest galaxies in the group), show any sign of group scale X-ray emission, we adapted the following sample selection strategy.

We draw galaxy groups from the 2PIGG catalogue
of galaxy groups \citep{b5}, based on the full release of
2dFGRS. We selected 2PIGG groups which satisfy the following criteria, 

I. Groups with at least five confirmed members ($N_{gal}\ge 5$): 
This provides us with a more reliable estimate of the group velocity
dispersion and reduces the danger of including groups of galaxies
which are not physical associated.

II. Groups with magnitude difference between first and second 
ranked galaxies $\Delta M_{12}\ge2.0$: This is the formal optical
criterion for fossil groups, and is imposed to ensure that the
$L^\star$ galaxies in groups have merged to form the central luminous
galaxy.

\begin{figure*}
  \begin{center}
    \leavevmode
      \epsfxsize=8cm\epsfbox{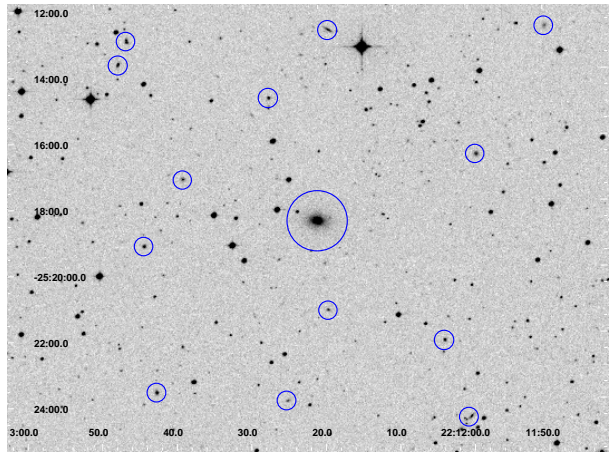}
      
       \caption[2515]{UKSTU Schmidt R-band image of the field of fossil candidate 
(2PIGG 2515) at $z\approx 0.062$ in 2dF south field. The circles 
show the group members with the dominant elliptical marked with larger 
circle at the centre of the field. The image is $12 \times 12$ arcmin$^2$. }
     \label{overlay}
  \end{center}
\end{figure*}

III. Groups with a giant elliptical galaxy as the BGG: The study of the 
volume-limited sample of fossils (Jones et al 2003) showed that the
brightest galaxy is always a giant elliptical. We therefore apply an
absolute magnitude cut of $M_B<-21.5$ and choose only elliptical
galaxies. Fig. \ref{overlay} shows an example group. 

Although the 2dF galaxy redshift survey, from which we selected our
targets, reaches $z\approx 0.2$, the optical criteria $\Delta
M_{12}\ge2.0$ and $M_{b_{J},BGG}\le -21.5$, limit the depth of the volume to 
$z<0.09$ within which the sample is complete. In addition, the $N_{gal}\ge5$ criterion 
requires at least four non-brightest galaxies to be spectroscopically confirmed members 
of these groups. While this criterion does not add any additional constraint on the volume, the 
number of galaxy groups will be substantially reduced. From the 2PIGG catalog and within $z<0.09$, 
we find that the trend in which the total number of galaxy groups decreases with increasing 
$N_{gal}$, is quite similar to the trend in which the number of groups with a large $\Delta M_{12}\geq2$ 
decreases with $N_{gal}$. However, the number of groups with $\Delta M_{12}\geq2$, for a given 
$N_{gal}$, is an order of magnitude smaller than the number of groups when this criterion is relaxed. 
This will be taken into account when we return to the estimation of the space density of fossil galaxy 
groups using this study, in section 5. 

An earlier volume limited sample, compiled during the WARPS project \citep{Scharf97}
was deeper and resulted in 5 fossils in a redshift range of
$z<0.24$ with a space number density of $4\times 10^{-6}h^3_{50}$
Mpc$^{-3}$ (8-20\% of all the systems with the same $L_X$). The 2dF
fields (north and south) cover a total area of $\sim1000$ deg$^2$
which is about 14 times larger than the area covered in the WARPS
X-ray survey, e.g. 73 deg$^2$. Hence we probe a significantly larger volume
in the present study than the X-ray selected WARPS sample. However,
the $N_{gal}\ge5$ selection criterion restricts us to richer
galaxy systems, and results in a statistically well-controlled sample
of 6 groups dominated by a luminous elliptical galaxy, judged based on the optical 
data available at the time. Four of these systems have been observed by \chandra, 
within the area covered by 2PIGG catalog and up to z=0.067.

As mentioned briefly in the introduction, the major distinction between this sample 
and that of Rasmussen et al (2006) are the two additional constraints, e.g. the brightest 
galaxy in the group is as luminous as a giant elliptical galaxy and the presence of a 
large magnitude gap such that the giant elliptical galaxy appears relatively isolated. 
We note that the photometric data based on which we selected our sample was shallow 
and thus the morphological information may not have been accurate. In light of the new and deeper 
photometric observations, as part of this study, some galaxy properties appear to be different. 
We will return to this point later in sections 3 and 5.

\begin{table*}
\begin{center}
\begin{tabular}{lcccccccc}
\hline\hline
2PIGG&RA      & Dec     & $z$ & M$_B$ &$\sigma_v$   &  $R_{vir}$ & $M_{halo, dyn}$ & $t_{obs}$ \\
ID & (J2000)   &  (J2000)    &     &       &  km/s  & $ kpc $ & $ 10^{12}h^{-1}M_{\odot} $ &  ks\\
\hline
1404 & 13 45 39.8  & -05 30 33 & 0.052 & -21.9 & 218    & 532 & 22.5   & 10 \\
1635 & 00 16 25.8  & -27 07 05 & 0.056 & -21.8 & 189   & 461 & 11.6   & 20 \\
2515 & 22 12 20.7  & -25 18 29 & 0.062 & -21.6 & 268  & 653 & 37.4   & 20 \\
2868 & 03 14 33.1  & -34 07 42 & 0.067 & -22.2 & 213    & 520 & 25.2   & 10 \\

\hline\hline
\end{tabular}
\caption{Properties of the 4 galaxy groups observed by \chandra. The coordinates refer to the 
position of the brightest galaxy in the group. The proposed exposure time is based on the expected fluxes to accumulate at least 1000 count per target using the $L_X-L_{opt}$ relation in \citet{kpj07} for fossil groups. Column (9) gives a dynamical estimate for the halo mass from the group velocity dispersion given in column (6). The sample is complete out to  $z=0.07$ noting that only groups with at least 5 spectroscopically identified members are included.}
\label{table1}
\end{center}
\end{table*}

\subsection{X-ray observations}

The main aim of the \chandra observations was to test for the
presence of extended thermal X-ray emission associated with a hot
intragroup medium, and to measure a mean gas temperature for all
targets. The four targets were centred with the brightest group galaxy
at the ACIS-S3 CCD aimpoint and observed in VFAINT mode for nominal
exposure times of 10--20~ks. These times were motivated by a desire to
accurately constrain the mean hot gas temperatures within
5--10~per~cent, based on the observed $L_X-L_B$ relation for X-ray
selected fossils \citep{kpj07}.

\subsection{Optical observations}

Optical observations of groups in this study were
performed during two observing runs in December 2007, May 2008 using the 2.5 meter
(100-inch) ${Ir\acute{e}\acute{n}ee}$ du Pont telescope, operating at Las
Campanas Observatory. The telescope was calibrated with the
wide field imager CCD (WFCCD) camera which covers a 25 arc-minute
diameter field with a scale of 0.77 arcsec/pixel.
The characteristic mean redshift  of groups is $z \sim 0.06$ which corresponds to a 
luminosity distance $D \sim 260$ Mpc and a scale of $1.13$ kpc/arcsec, 
based on our cosmological assumptions.

For each galaxy group a total exposure time of 3600s and 1440s were applied in B-band and
R-band respectively. Since each group image is a mosaic frame consisting of four individual
pointing, the actual exposure time used for each group is one fourth of those values, i.e. 900s
in B-band and 360s in R-band. To avoid pixel saturation due to bright sources and be able to 
remove cosmic rays as well as increasing the signal to noise, 
exposures were split into 300s and 120s in B-band and R-band, respectively.  
The photometric standards observations were also carried out at the time of the run with exposure 
times of 10 to 16 seconds in B-band and 4 to 8 seconds in R-band.

\section{X-ray analysis}

The {\em Chandra} data were re-calibrated and analysed using {\sc
  ciao} v4.5 with CALDB v4.5.7. Bad pixels were screened out using the
bad pixel map provided by the pipeline. Standard cleaning of
background flares revealed no periods of significantly enhanced
background in any of the data sets. Point sources were identified with
the {\sc ciao} task 'wavdetect' and masked out in all subsequent
analysis.

\begin{figure}
  \begin{center}
    \leavevmode    
\epsfxsize=8.5cm  
\epsfbox{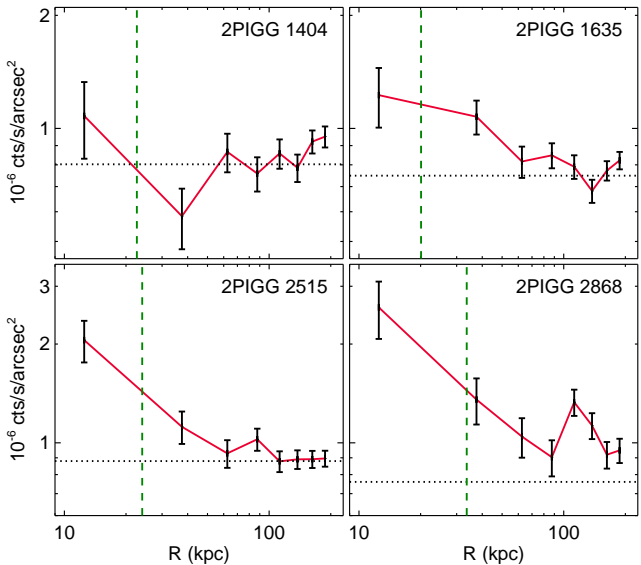} 
\caption[SB]{{\em Chandra} 0.3--2~keV X-ray surface brightness
  profiles of the groups. The horizontal dotted lines show the background level
  evaluated from a surrounding annulus in the source data. {\bf The vertical dashed 
  lines mark the isophotal radius reaching 25 mag arcsec$^{-2}$ in B-band, e.g. 
  0.5\,D$_{25}$, for the brightest group galaxy. The X-ray surface brightness extends 
  well beyond the optical size (D$_{25}$) of the most luminous galaxy in the group with 
  the exception of 2PIGG 1404.} 
   }
\label{SB}
  \end{center}
\end{figure}

\begin{figure*}
  \begin{center}
    \leavevmode
    
     \epsfxsize=8.0cm  \epsfbox{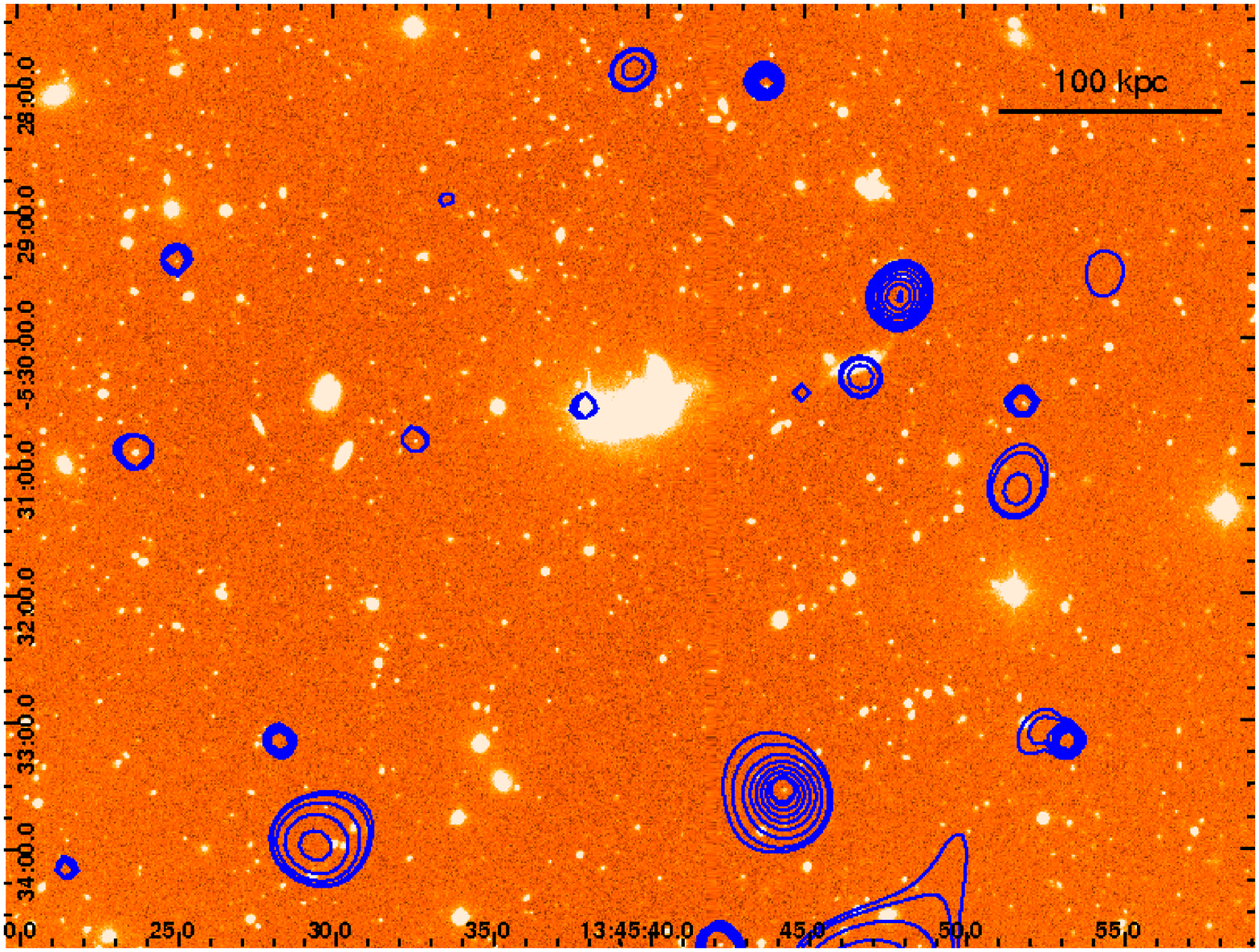} %1404 ds9-R4.eps
     \epsfxsize=8.0cm  \epsfbox{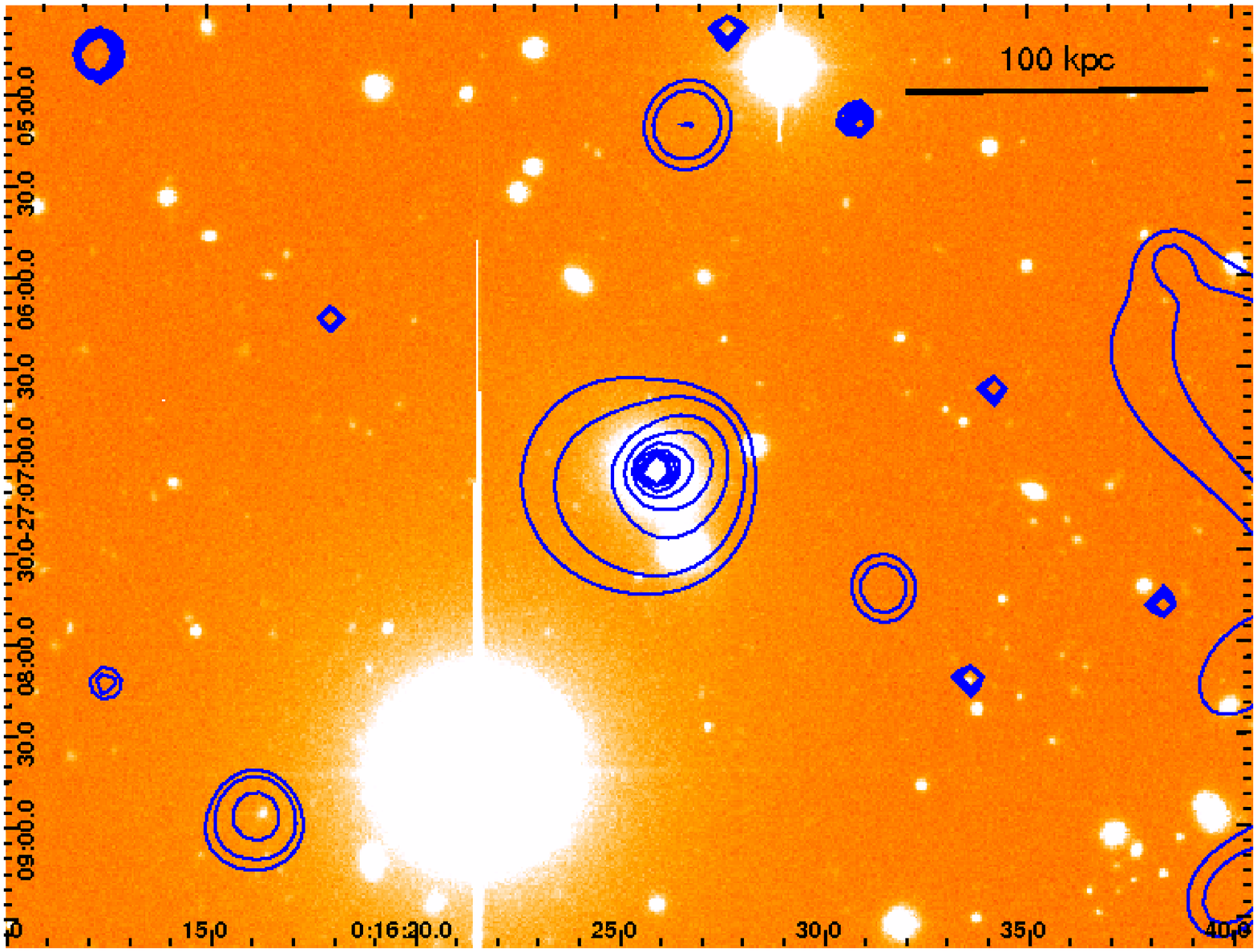} %1635
     \epsfxsize=8.0cm  \epsfbox{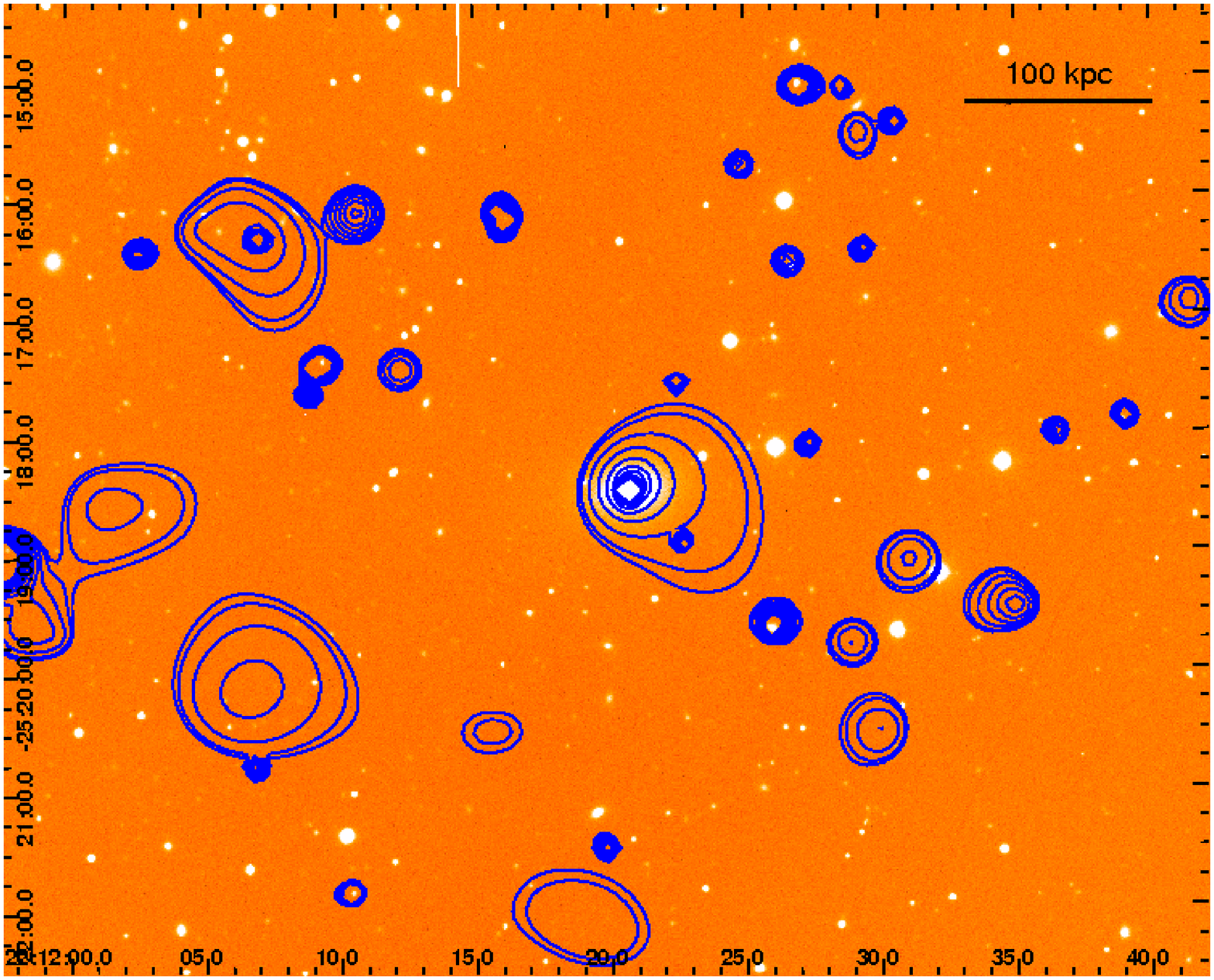}  %2515
     \epsfxsize=8.0cm \epsfbox{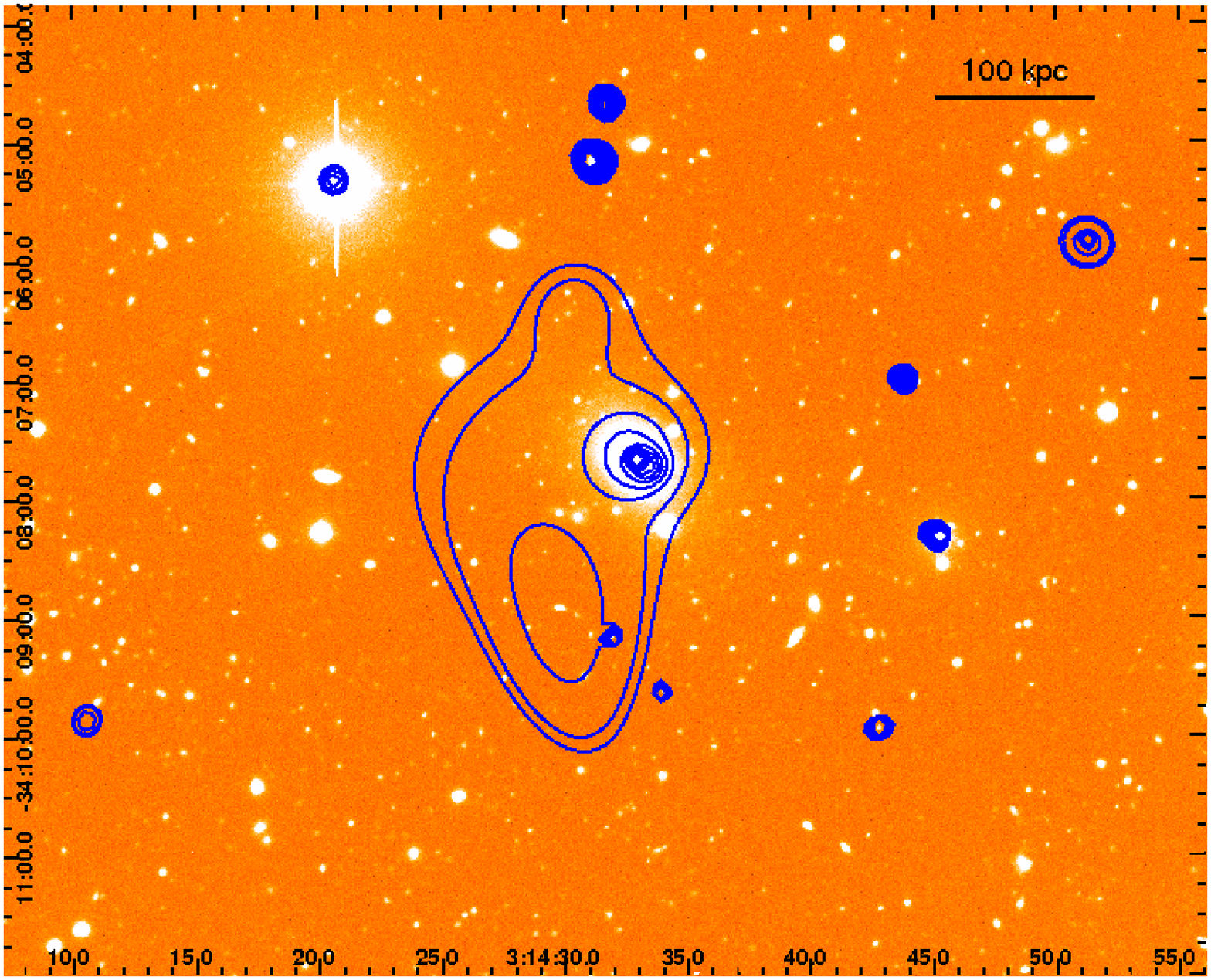} %2868

\caption[2515]{The smoothed X-ray contours over optical R-band images of the central regions. The images are 2PIGG 1404, 1635, 2515 and 2868 from left to right and top to bottom, respectively. Three out of the four groups show signatures of extended X-ray emission. See Fig. \ref{zoom} for the zoomed in optical image of the core of  2PIGG 1404 for which we find no extended X-ray emission.}
\label{overlay2}
  \end{center}
\end{figure*}

To search for the presence of extended X-ray emission,
exposure-corrected surface brightness profiles were extracted from the
BGG optical centre for each group. Due to the anticipated limited
spatial extent and soft spectral nature of the sources, the background
level was evaluated using a surrounding on-chip annulus. This ensures
a reliable estimate of the local soft X-ray background. The resulting
profiles are shown in Fig. \ref{SB}. Despite displaying $\la
200$~net counts each, three of the groups (2PIGG\,1635, 2515 and 2868)
show clear evidence ($>3\sigma$ significance) for extended emission,
with only 2PIGG\,1404 remaining X-ray undetected. The detected
emission extends to projected radii of $\ga 50$--100~kpc ($\ga
0.8'-1.5'$) in all three cases. This is well beyond the width of the
{\em Chandra} point spread function as well as the BGG optical extent,
strongly suggesting the presence of intragroup emission. {\bf In Fig. \ref{SB}
we show the optical extend of the BGGs quantified using $D_{25}$, e.g. the
isophotal diameter at the surface brightness of 25 mag arcsec$^{-2}$ in B-band, 
with values given in Table 3. Except for the 2PIGG\,1404, the X-ray surface brightness 
extends beyond the corresponding isophotal radius.} 
Fig. \ref{overlay2}, which shows $R$-band optical images of our four groups 
(cf.\ Section~2.2) with smoothed X-ray contours overlayed.

Although all three X-ray detected groups are rather X-ray faint,
useful constraints on the spectral nature of their X-ray emission
could still be obtained. The spectral analysis assumed an APEC hot
plasma emission model at the source redshift, absorbed by the Galactic
H{\sc i} column density at the relevant sky position \citep{kalb05}.
Given the limited photon statistics, the plasma abundance was fixed at
an assumed value of 0.4~Z$_\odot$ in all cases. Spectra were extracted
within $R = 100$~kpc apertures ($R=200$~kpc for 2PIGG\,2868,
cf.\ Fig. \ref{SB}), accumulated in bins of $\geq 20$~counts, and
fitted in {\sc xspec} v12.8 assuming $\chi^2$ statistics. Again,
surrounding on-chip annuli were used for background estimation. In all
cases, an APEC model was found to provide a better fit than a
power-law, thus testifying to the thermal nature of the emission. To
test the robustness of the spectral fitting for these low-surface
brightness systems, we also obtained spectra in bins of just $\geq
5$~net counts and fitted them assuming Cash statistics. In addition,
we assumed an APEC spectrum and inferred X-ray temperatures from
imaging data alone, using exposure-corrected (0.5--1)/(1--2)~keV
hardness ratios. Encouragingly, all three methods produced consistent
temperature estimates for each of the three groups, lending
credibility to the results.

Table~\ref{table2} summarises the observational characteristics and
derived X-ray properties for all four targets, including $1\sigma$
uncertainties on spectral parameters. For the X-ray undetected
2PIGG\,1404, a $3\sigma$ upper limit on $L_{\rm X}$ is quoted,
obtained from its count rate Poisson error and assuming
$T=1.0$~keV. In summary, the imaging and spectral analyses present a
coherent picture in which three of the four target groups display
extended, $T \approx 1$--2~keV thermal emission with bolometric
$L_{\rm X,bol}$ in the range (0.3--2)$ \times
10^{42}$~erg~s$^{-1}$. To allow straightforward comparison to the
\citet{jones03} X-ray criterion for fossils, we also extrapolated the
inferred luminosities out to $0.5R_{\rm vir}$, taking $R_{\rm vir}$
from Table~\ref{table1}. For a standard $\beta$--model for the surface
brightness with $\beta=0.5$ and $r_c=20$~kpc, this would increase the
luminosities quoted in Table~\ref{table2} by a modest 10--70~per~cent.

\begin{table*}
\begin{center}
\begin{tabular}{ccccccccccc}
\hline\hline
2PIGG  & $t_{\rm exp}$ & $R$   & counts & S/N & $N_{\rm H}$ & $T$ & $Z$ & $L_{\rm X, 0.3-2\,keV}$ & $L_{\rm X,bol}$ &$L_{\rm X,bol}(\le r_{500})$\\
 ID    & (ks) & (kpc) &  & & ($10^{20}$~cm$^{-2}$) & (keV) & (Z$_\odot$) & ($10^{41}$~erg~s$^{-1}$)& ($10^{41}$~erg~s$^{-1}$)&($10^{41}$~erg~s$^{-1}$) \\ \hline
1404 & 10.7 & 100 & $-6$ & 0.0 & 1.84 & $1.0^\ast$         & $0.4^\ast$ &  
   $<1.3$ & $<2.1$ & $<2.1$ \\
1635 & 19.4 & 100 & 69   & 3.5 & 3.04 & $1.6^{+1.7}_{-0.6}$ & $0.4^\ast$ &  
   $1.4 \pm 0.2$ & $2.5 \pm 0.3$&$4.7\pm0.6$\\
2515 & 19.9 & 100 & 73   & 3.7 & 1.99 & $1.1^{+0.3}_{-0.2}$ & $0.4^\ast$ &  
   $1.7 \pm 0.2$ & $2.7 \pm 0.3$&$4.7\pm0.5$\\
2868 &  9.9 & 200 & 197  & 7.6 & 1.95 & $2.2^{+1.1}_{-0.4}$ & $0.4^\ast$ &  
   $10.1 \pm 1.5$ & $19.1 \pm 2.9$&$26.9\pm4.1$\\
\hline%\hline
\end{tabular}
\caption{Derived X-ray properties of the target groups: Cleaned
  exposure times $t_{\rm exp}$, radius $R$ of the circular aperture
  used for extraction of X-ray properties, 0.3--2~keV net counts and
  signal-to-noise (S/N) ratios, Galactic absorbing column $N_{\rm H}$,
  and spectral parameters including pseudo-bolometric (0.05--10~keV)
  luminosities. Asterisk indicates an assumed value.}
\label{table2}
\end{center}
\end{table*}

\section{Optical properties}

While the primary aim of the deep optical observations was to obtain a more accurate view of the groups and the brightest group galaxy, we also needed to obtain total optical luminosity of the galaxies in groups through photometric means. As a byproduct of this analysis we also obtain the galaxy luminosity function for future studies.

\subsection{Star-galaxy separation}

The imaging data for all 4 groups were reduced using IRAF Packages. Objects in all images were detected by  SExtractor \citep{b1}. Each detected object in the image catalogs is associated with a FLAG value which shows the degree of reliability of the measurements. After some iteration, we included all catalog entries with FLAG $ {\leq3}$, given the quality of the photometric data and our desire to not exclude genuine galaxies. Catalogue entries with FLAG $>3$ constitute to only $\sim 2 $ per cent of the total number of entries. 

Each entry in the catalogue also includes a stellarity class parameter, STAR, with values from 0 to 1. Lower the value, the more likely the detected object is an extended source, e.g. galaxy. The images and the radial profile of a number of objects with different values of STAR parameter were examined in IRAF before deciding the borderline. Typically, objects with STAR ${>0.83}$ were found to be stars, whilst those with ${\le0.83}$ were galaxies.

\subsection{Photometric completeness }
\begin{figure*} 
   \begin{center}
     \leavevmode
       \epsfxsize=8cm\epsfbox{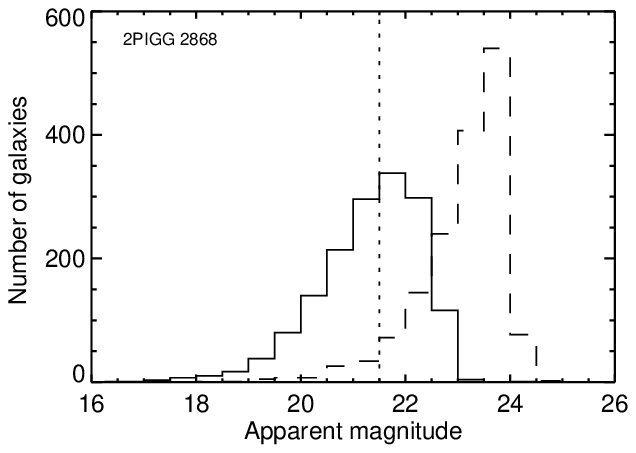}
       \epsfxsize=8cm\epsfbox{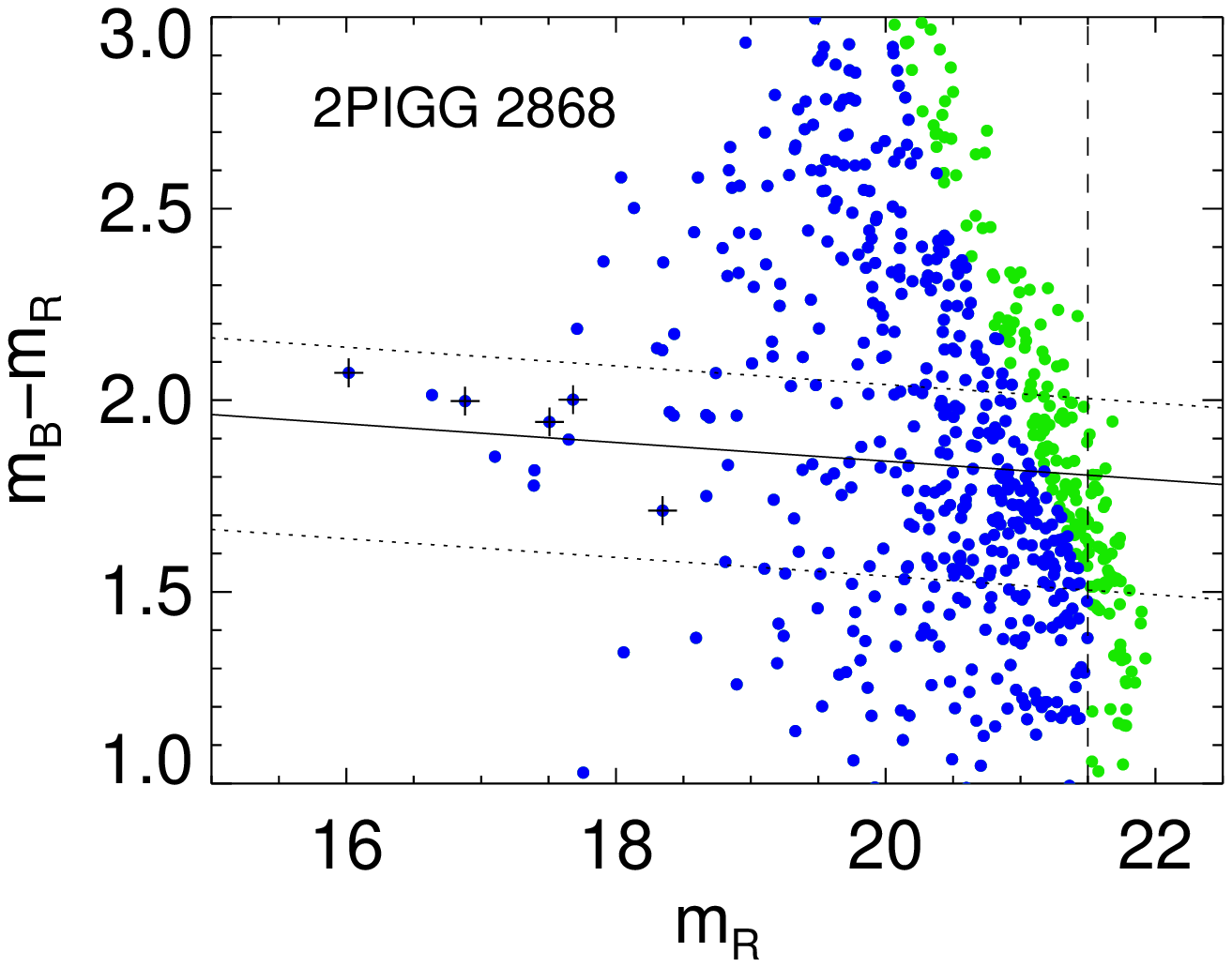}
\caption[color]{Left: An example magnitude histogram of galaxies for 2PIGG 2868 in the R-band (solid) and the B-band (dashed). Right: Colour magnitude diagram of galaxies in the observed fields of view of 2PIGG 2868. The crosses mark the spectroscopically confirmed group members. The dark/blue filled circles show galaxies within the field of view within the applied R- band and B- band magnitude completeness limit. The gray/green filled data points mark galaxies in the observed field of view but outside the magnitude completeness limit.  The solid black line is a linear fit to all spectroscopically confirmed group members while the upper and lower lines are the $2\sigma $ confidence levels (dotted lines), within which the galaxies are assumed to be the group members. }
      \label{completeness}
   \end{center}
 \end{figure*}

As an example, the completeness level of the B- and R-band photometric data for 2PIGG 2868 is shown in Fig. \ref{completeness}. The solid and the dashed histograms show the magnitude distribution of the galaxies in the field in R- and B-band, respectively. The adopted magnitude limit in the R-band is shown by a vertical dashed line. This limit was determined considering the position of the peak in the apparent magnitude histogram for each group.
 
 \subsection{Colour-magnitude diagram and group membership}
In the absence of spectroscopic redshift measurements, the colour magnitude diagram (CMD) 
is often used to assign galaxies to groups, in particular since early-type galaxies form a red 
sequence which is a characteristic of their evolution \citep{b3}.

The total apparent magnitude and magnitude within an aperture of 5 arcsec were obtained 
in both R- and B-band filters. The ${m_{\rmn{B}}-m_{\rmn{R}}}$ colour for all objects identified 
as galaxies within a virial radius is presented against the total apparent magnitude in Fig. \ref{completeness}. 
The spectroscopic members based on the redshift of galaxies around the BGG within the 
virial radius of groups were also found using the NED data base and the 2dF galaxy redshift survey. 
Galaxies were assigned as spectroscopic group members if they fall within a redshift range of $0.005\times (1+z_{g})$ from BGG redshift \citep{b8}.

As an example, Fig. \ref{completeness} shows the CMD of 2PIGG 2868, with crosses representing spectroscopic group members. The blue points are all galaxies within field of view after applying the R- and B-band magnitude completeness. The solid line is a linear fit to all spectroscopic early type galaxy members while the two lower and upper lines are ${\pm 2 \sigma }$  confidence levels. 

\subsection{Galaxy luminosity function}
The study of the luminosity function of galaxies in different environments provide
important observational constraints for cosmology and galaxy formation and evolution models. Schechter (1976) and Turner \& Gott (1976) determined the galaxy luminosity functions of groups and rich clusters. Schechter (1976) also proposed the following form
for the luminosity function according to the observed data of clusters and field galaxies 
 \[
 \Phi(M)dM = (0.4 ln 10)\Phi^{*}X^{(1+\alpha)}\times exp^{-X}dM   \hspace{16mm}  (1)
 \]
 where $X=10^{0.4(M^*-M )}$ and the quantity $\Phi(M)$ is proportional to the number of galaxies that have
 absolute magnitudes in the range (M, M+dM), $\Phi^{*}$ is the characteristic number density,
  $M^{*}$ is the characteristic absolute magnitude. 
  
At the bright end, the Schechter function drops sharply. It rises at the faint-end following a power law with a slope given by $\alpha$. Thereby, the faint-end slope is decreasing, increasing, or flat for $\alpha > -1$, $\alpha < -1$, and $\alpha = -1$ respectively.

Galaxies were selected as group members on the basis of the estimated upper and lower
 confidence levels found in previous section from a linear fit to ${m_{\rmn{B}}-m_{\rmn{R}}}$ colour of groups members. Therefore, a ${m_{\rmn{B}}-m_{\rmn{R}}}$ colour cut-off was applied
 at ${m_{\rmn{B}}-m_{\rmn{R}}}{=(-0.023 \pm 0.002)\times m_{\rmn{R}}}{+(2.327\pm0.034)}$
 where $ m_{\rmn{R}}$ is the apparent R-band magnitude of any object in the fields. 

In addition, a statistical method of background subtraction was employed to remove the contamination from
foreground-background objects. To obtain a background galaxy sample, regions beyond $r_{200}$ radius in each group mosaic image were selected as a background field, roughly corresponding to an area of 
1 $deg^2$. For each group, the net galaxy luminosity function was determined by subtracting the normalised
background luminosity function, from the luminosity function for the same group in the R-band. 

\begin{figure*}
  \begin{center}
    \leavevmode
      \epsfxsize=8cm\epsfbox{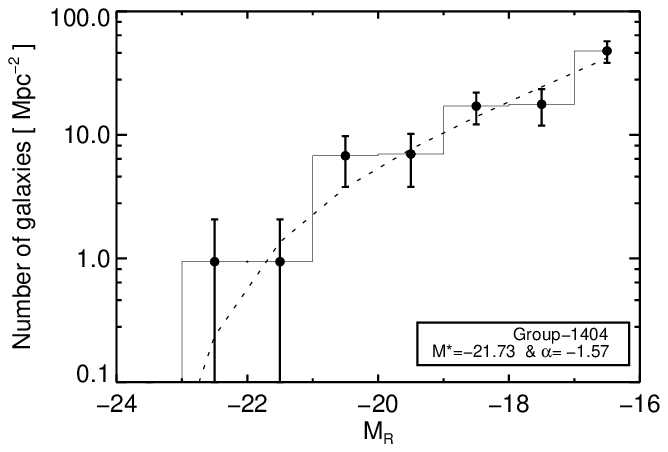}
      \epsfxsize=8cm\epsfbox{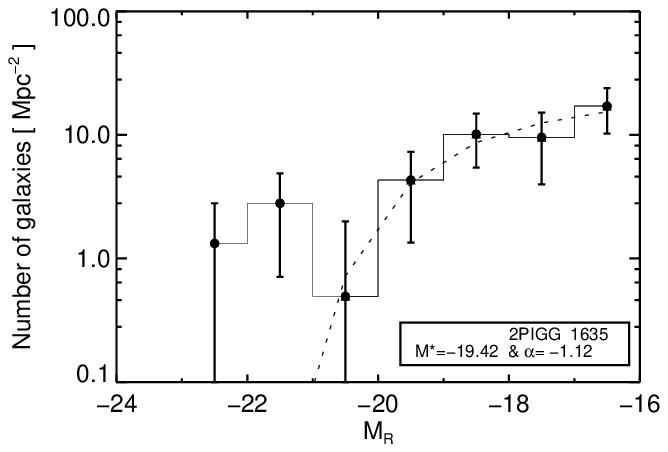}
      \epsfxsize=8cm\epsfbox{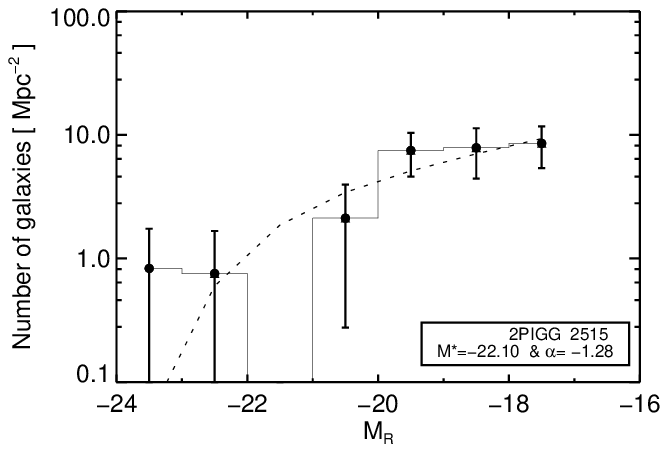}
      \epsfxsize=8cm\epsfbox{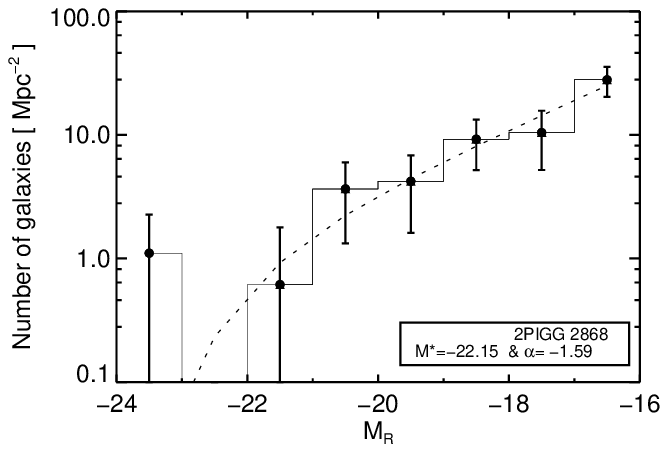}
       \caption[2515]{The galaxy luminosity function for the sample groups,  2PIGG 1404, 1635, 2515 and 2868 from left to right and top to bottom. The fitted Schechter parameters are also given for each luminosity function.}
     \label{LF}
  \end{center}
\end{figure*}

The group R-band luminosity functions are shown in Fig. \ref{LF} for individual groups. The curve shows the 
single Schecther function fit on the luminosity function of group members.
We excluded the BGG in the LF fitting, because the large luminosity gap results in a poor 
fit and poor constraint for $M^{*}$ values. A single Schechter function of the form of Eq. 1. 
yielding the best fitting values in R-band are given in Table~\ref{table3}.

We calculate the dwarf to giant ratio of our sample groups using their luminosity functions. We assume galaxies brighter than -19.0 mag as giants and dwarf galaxies, with absolute magnitude between -19.0 and -17.0 (or -16.0). Results for dwarf to giant ratio are presented in Table \ref{table3}.
 
\begin{table*}
\begin{center}
\begin{tabular}{lccccccccc}
\hline\hline
2PIGG  ID &  M$_{B}^{BGG}$ & M$_{R}^{BGG}$&B-R$^{BGG}$& $\Delta m_{12}^{0.5vir}$ &D$_{25}^{BGG}$(kpc)& M$_{R}^{*} $  & $\alpha$ & $\frac{N_{Dwarf}}{N_{Giant}}^{I}$ & $\frac{N_{Dwarf}}{N_{Giant}}^{II}$\\
\hline
1404   & -21.0 & -22.6 & $1.8 \pm 0.1$ & 1.6 &45.2 & $-21.73 \pm 0.46$ &$ -1.57 \pm 0.06$ & 2.2 & 5.3 \\
1635   & -20.8 & -22.7 & $2.0 \pm 0.1$ & 1.1 &40.2 & $-19.42 \pm 0.36$ & $-1.12 \pm 0.10$ & 2.2 & 4.1\\
2515   & -21.3 & -23.4 & $2.1 \pm 0.1$ & 3.4 &47.9& $-22.10 \pm 0.52$ & $-1.28 \pm 0.06$ & 1.5 & $-$\\
2868   & -21.2 & -23.1  & $2.0 \pm 0.1$ & 2.5 &67.3& $-22.15 \pm 0.48 $& $-1.59 \pm 0.06 $ & 2.1 & 5.0\\

\hline\hline
\end{tabular}
\caption{Optical properties of the sample groups. Column (2) to (4) are our measured BGG magnitudes in B- and R-bands and B-R colour, respectively. {\bf Column (5) gives the diameter, D$_{25}$, of the BGG in kpc reaching the surface brightness of 25 mag/arcsec$^2$} in B-band. Column (6) and (7) are the R-band galaxy luminosity function parameters from a single Schechter fit. Dwarf to giant ratios are also given in the last two columns, (I) Giants: $M_{R}\leqslant -19$ and Dwarfs: $-19\leqslant M_{R} \leqslant-17$), (II) Giants: $M_{R}\leqslant -19$ and Dwarfs:$-19\leqslant M_{R} \leqslant-16$). }
\label{table3}
\end{center}
\end{table*} 

Two of the sample groups, 2PIGG 2515 and 2868, show a large magnitude gap ($\ge 2$ mag) between the second brightest galaxy within the half of the Virial radius and the BGG, in R-band. Although 2PIGG 1404 shows a magnitude gap of 1.6 mag, the BGG shows a disturbed morphology indicating that is in the process of merging with the second brightest galaxy and futures associated with on going merger are present. {\bf The BGG in this case is not a giant elliptical galaxy.}

At the core of the group 2PIGG 1635 there is a bright galaxy close to the BGG with about 1 magnitude gap. In the absence of sufficient information to rule out projection effect, we assume that these two galaxies are in the pre-merger state. The reason this group was included in the candidate sample, in the first place, was poor photometric measurement based on which the 2dF galaxy redshift survey was built, at the distance of this group. 

\begin{figure*}
  \begin{center}
    \leavevmode
      \epsfxsize=8cm\epsfbox{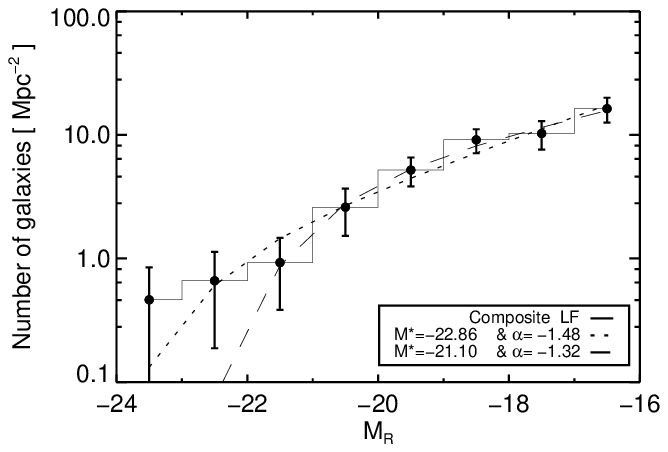}
      \epsfxsize=8cm\epsfbox{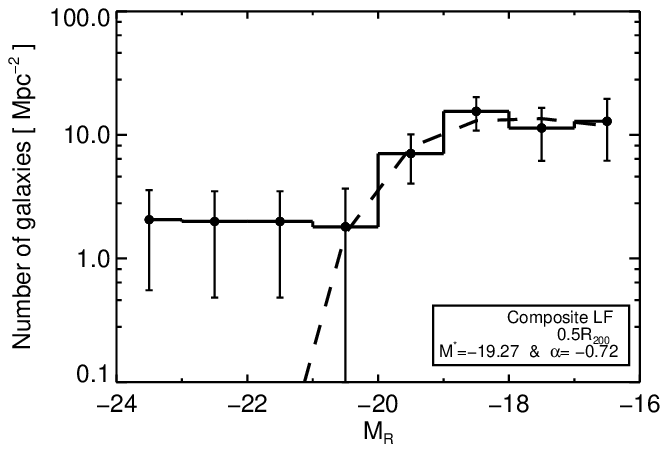}

\caption[lfcomposite]{The composite luminosity function of four groups in our sample, within the $r_{200}$ radius (left) and 0.5 $r_{200}$ (right). The dotted and dashed curves indicate the Schechter function fits to the observed composite luminosity function including and excluding the BGGs, respectively.}
     \label{lfcomposite}
  \end{center}
\end{figure*}

We also derived the composite luminosity function of the four galaxy group in R-band within Virial radius and half the Virial radius.  The composite luminosity function has been constructed by combining the individual luminosity 
functions of four galaxy groups. Number of galaxies in each magnitude bin have been normalised to the total
area of groups contributing in corresponded bin. Fig. \ref{lfcomposite} shows the composite luminosity function within $r_{200}$ (left panel) and 0.5$r_{200}$ (right panel). The dotted and dashed curves (left panel in Fig. \ref{LF}) indicate the commonly-used Schechter function fits on the observed composite luminosity function including and excluding the BGGs, respectively.  The best values of Schechter parameters in each fit have been computed as $M^* = 21.10 \pm 0.18$ (without BGGs), $M^* = 22.86\pm0.24$ (with BGGs),  $\alpha = 1.32 \pm 0.02$ (without BGGs) and $\alpha = 1.48 \pm 0.03$ (with BGGs). The best Schechter function fit to the observed composite luminosity function of groups within half of the Virial radius (right panel in Fig. \ref{LF}) with excluding the BGGs gives Schechter parameters as $M^{*} =19.27 \pm 0.14$ and $\alpha = 0.72 \pm 0.07$.

\subsection{A brief note on 2PIGG 1404}

\begin{figure*}
  \begin{center}
    \leavevmode
      \epsfxsize=8cm\epsfbox{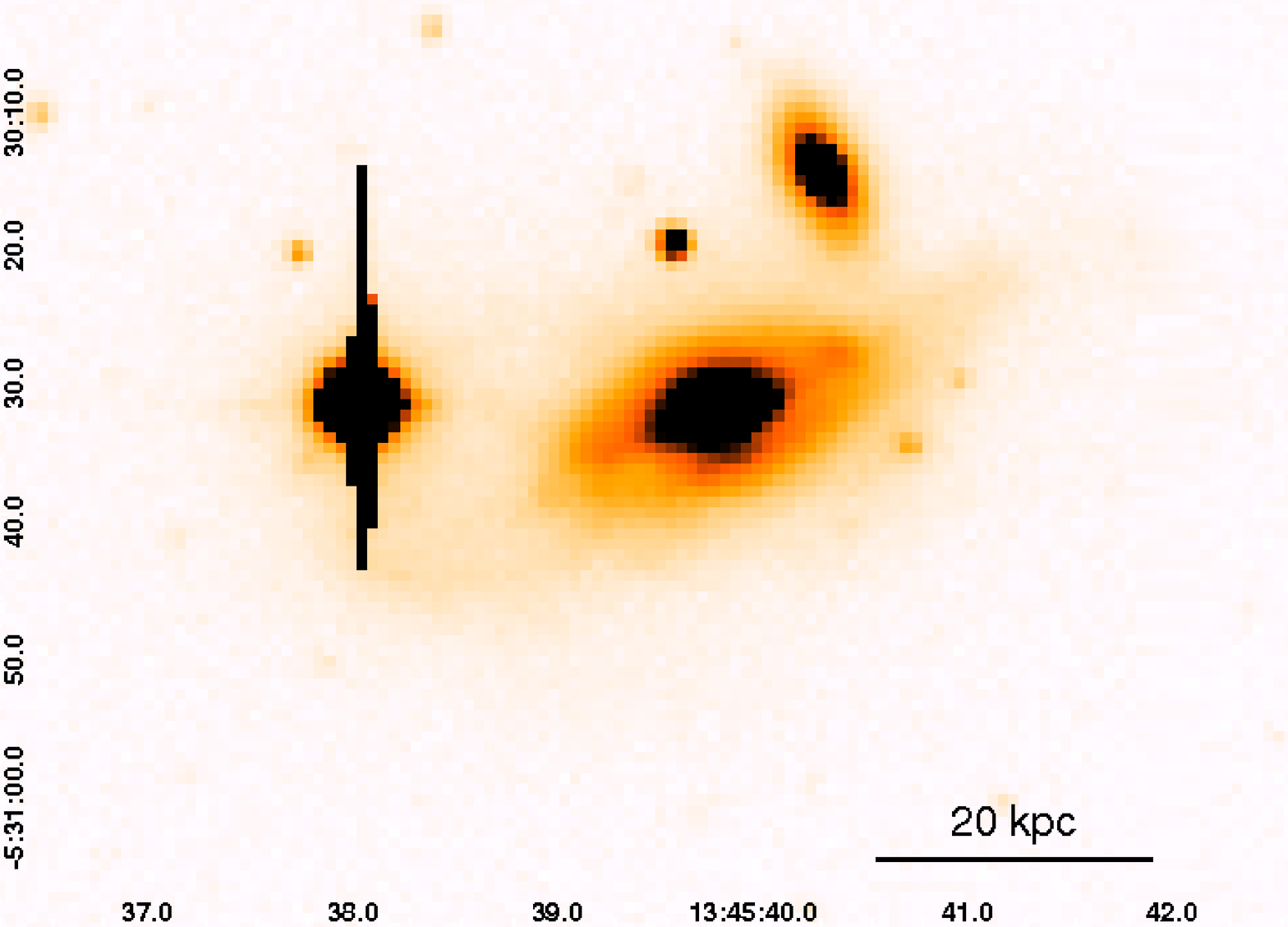}
\caption[zoom]{A zoomed in R-band image of the 2PIGG 1404 showing a disturbed morphology of the brightest group galaxy indicating an ongoing tidal interaction with its nearest galaxy.}
     \label{zoom}
  \end{center}
\end{figure*}

The brightest galaxies in all of the groups are giant elliptical galaxies except for the brightest galaxy in  2PIGG1404, which appears to exhibit disk like structure. This galaxy is slightly bluer ($\sim0.2$ mag) than the other brightest galaxies in the groups in our sample. A closer look at the surrounding of the BGG reveals a tidal interaction between this galaxy and its companion galaxy (Fig. \ref{zoom}). There is also a foreground stellar source which matches the location of the detected X-ray point source (Fig. \ref{overlay2}, top-left). {\bf Given the nature of this group in the optical, we argue that this galaxy group should not have been in our sample in the first place as the BGG morphology does not satisfy the initial selection criteria (condition II and III, section 2). As a result the major contrast between this galaxy group and the other groups in our sample is the BGG morphology.}

\section{Discussion and conclusions}

We carried out \chandra X-ray observations of a sample of 4 galaxy groups, each {\em appeared} to be dominated by a giant elliptical galaxy in $b_{J}$-band images, to explore whether they are also associated with a group scale X-ray emissions. If detected, one could provide a direct evidence that luminous elliptical galaxies form within the collapsed core of galaxy groups where the ionised gas reaches a $\sim$1 keV temperature. The X-ray emission itself can be used to probe the gravitational potential of the group which is complementary to the dynamical mass estimation, often used for galaxy groups. 

The X-ray analysis shows that, although the detected X-ray emission is primarily associated with the brightest group galaxies, there are clear evidences for the extended X-ray emission when the BGG is {\em genuinely} a giant elliptical galaxy. Our deep optical observations, following the X-ray observations, show that one of the BGGs is not a giant elliptical galaxy, despite what we had planned based on the optical observations available at the time and prior to the X-ray observations. We found 3 galaxy groups with giant elliptical BGGs to be associated with X-ray emission extending beyond the optical size of the BGG. One of these 3 galaxy groups (2PIG 2868) meets the requirements for fossil groups, satisfying both the X-ray luminosity and the optical luminosity gap criteria, while two remaining galaxy groups marginally satisfy the X-ray luminosity within the measurement uncertainty, of which only one meets the optical criterion on the luminosity gap, noting that for the cosmology adapted in this study the conventional threshold for the X-ray luminosity is $0.5\times10^{42} h_{70}^{-2}{\rm erg\,s}^{-1}$.

\begin{figure*}
  \begin{center}
    \leavevmode
      \epsfxsize=8cm\epsfbox{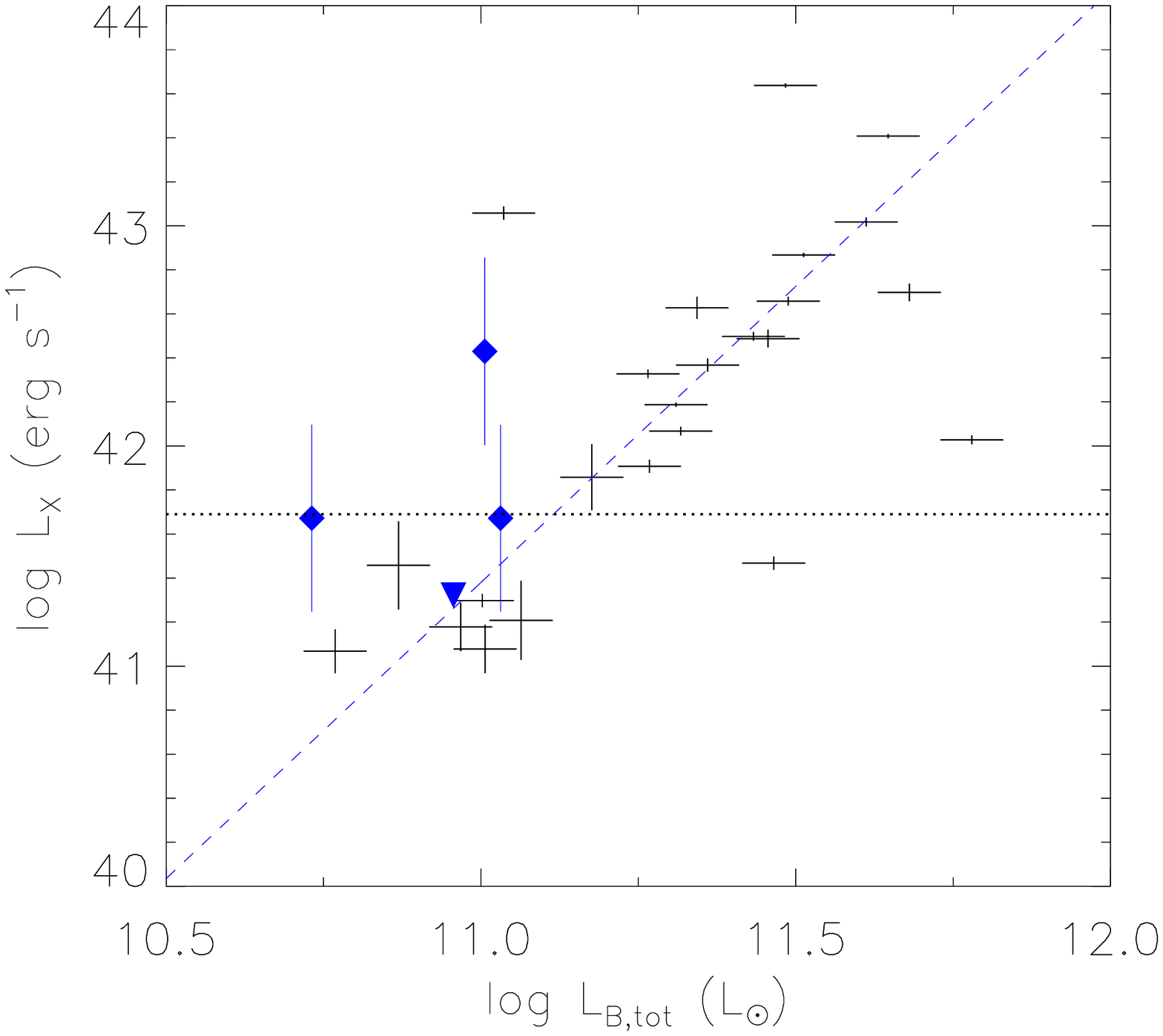}
      \epsfxsize=8cm\epsfbox{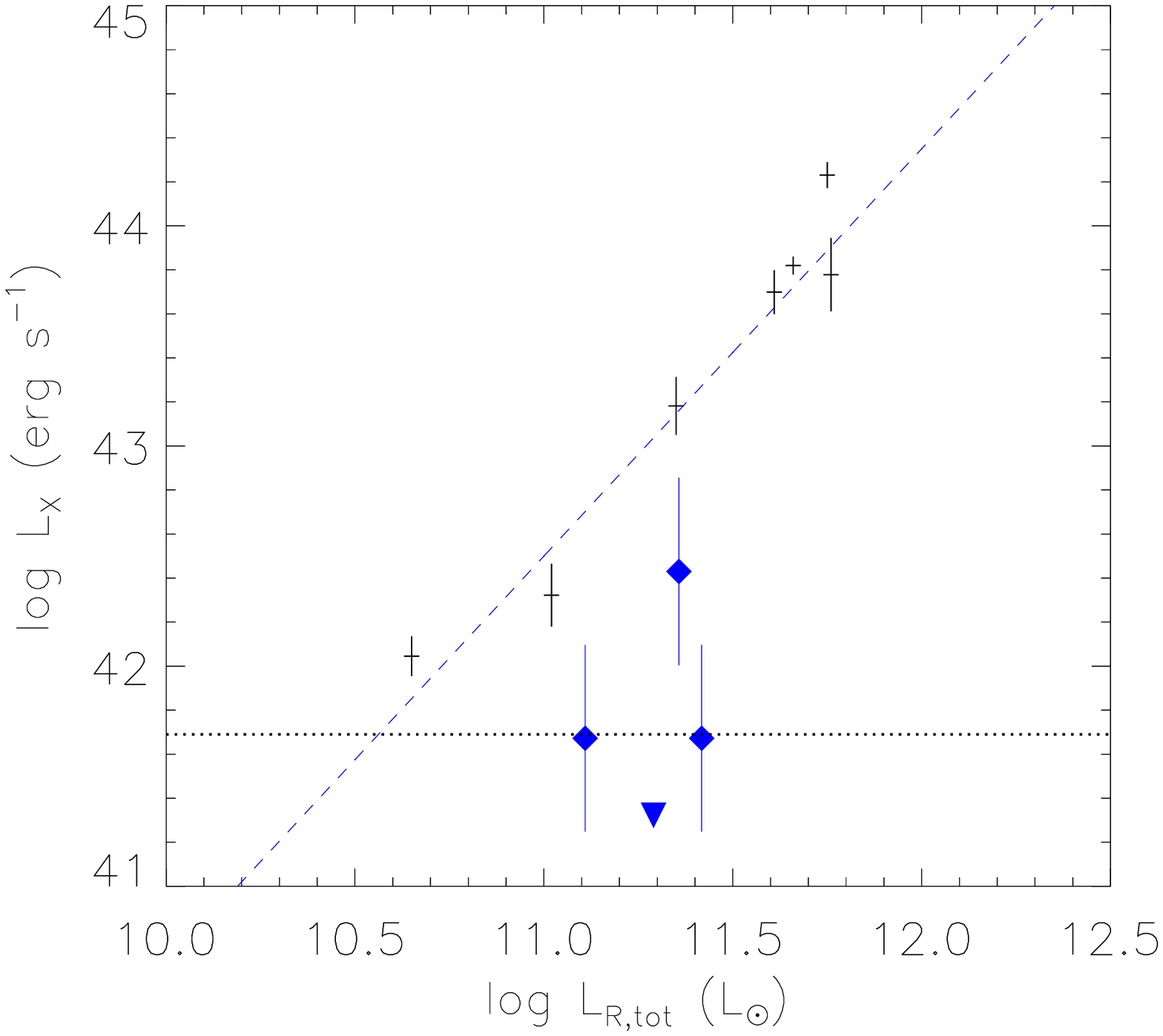}
      
\caption[sr]{The $L_{opt}:L_X$ relation for the optically selected groups in this study (dark/blue symbols) and the X-ray bright groups \citep{helsdon03} (left) and the X-ray selected fossil groups \citep{kpj07} (right). The $L_{B,tot}$ and $L_{R,tot}$ refer to the total B- and R-band luminosity of the group.The fitted dashed line is the best fit to the presented comparison data. The dark/blue triangle represents 2PIGG 1404 for which only an upper limit X-ray flux is available.}  
     \label{sr}
  \end{center}
\end{figure*}

\begin{figure*}
  \begin{center}
    \leavevmode
     \epsfxsize=10cm\epsfbox{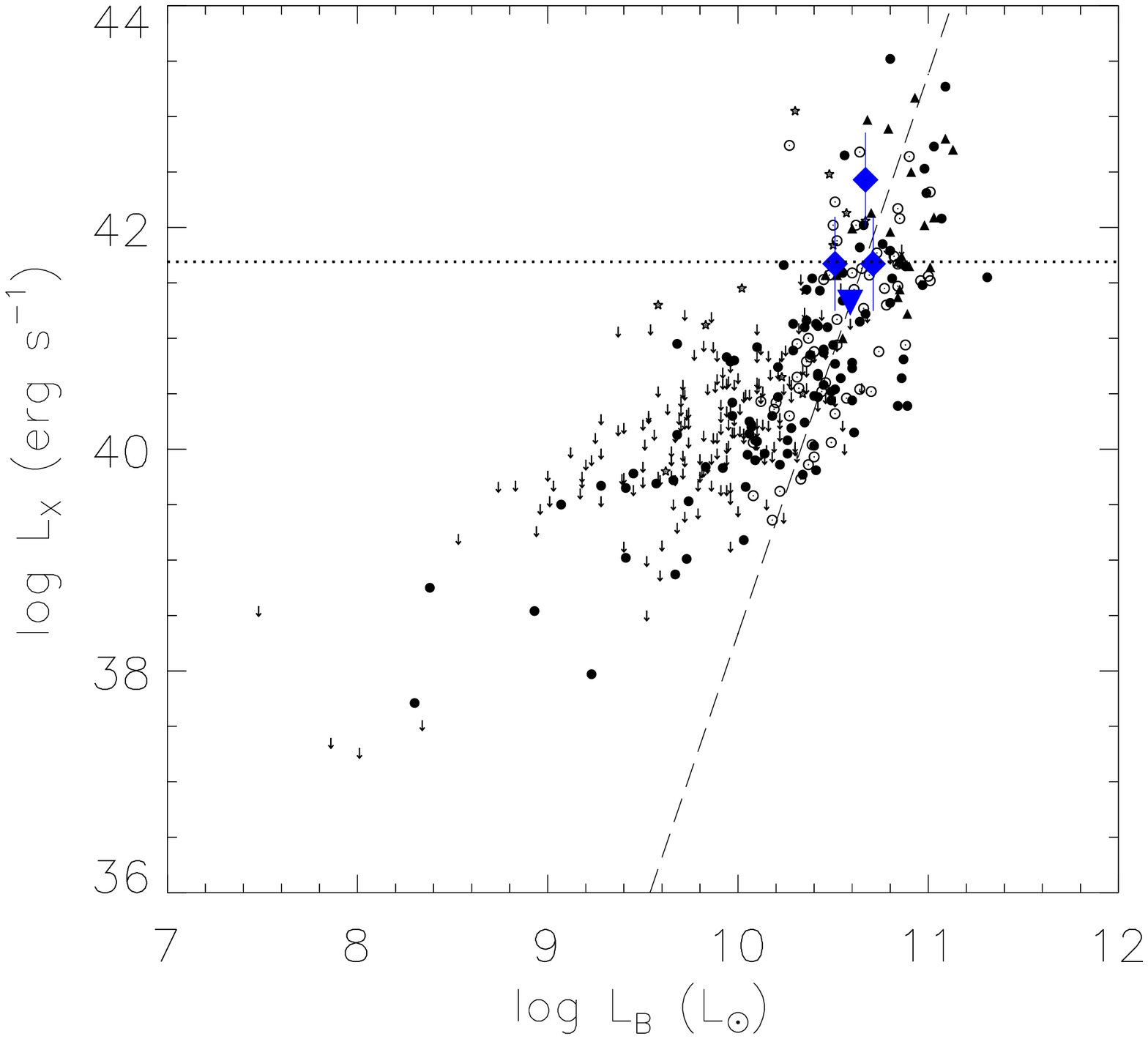}

\caption[sr]{A comparison between brightest group galaxies in our sample and the various early type galaxies in \cite{ellis06} in the plane of $L_X:L_B$. The L$_{B}$ refers to the B-band luminosity of the galaxies. The symbols are the same as in their Fig. \ref{SB}, triangles for BCGs, open circles for BGGs, stars for AGN, arrows indicating the upper limits for any type and filled circles are other types of galaxies.The dashed line is the best fit for the $L_X:L_B$ associated with the BGGs in their sample. The X-ray luminosity of the groups for which we detect extended X-ray emission, all except 2PIGG 1404, have been extrapolated to their $r_{500}$. The horizontal dotted line marks the fossil X-ray threshold as given by \cite{jones03}. Three of the groups show $L_X$ above or close to this threshold.}
     \label{xsr}
  \end{center}
\end{figure*}

In Fig. \ref{sr} we present $L_X:L_{opt}$ scaling relations for groups and galaxy sample in order to compare with our observations. The $L_{B,tot}$ and $L_{R,tot}$ refer to the group B- and R-band luminosity in solar luminosity units. As seen in this figure, the groups studied here follow the general trend of X-ray bright groups in \cite{helsdon03} but are dimmer in X-ray compared to the X-ray selected fossil groups in \cite{kpj07}. Fig. \ref{xsr} compares the X-ray luminosity associated with early type galaxies as a function of their B-band optical luminosity, L$_{B}$, between \citet{ellis06} sample and this study. Our sample galaxies are generally as X-ray luminous as the BGGs in other groups, for their optical luminosity. 

The X-ray luminosity of the 2PIGG 2868 is $L_X=1.9\times 10^{42}{\rm erg\,s}^{-1}$ which clearly satisfies the X-ray criterion for fossils. Assuming a beta model for the X-ray surface brightness distribution with $\beta=0.5$ and core radius, $r_c=20$ kpc, the extrapolated X-ray luminosity will be $L_X(500)=2.7\times10^{42}{\rm erg\,s}^{-1}$, within $r_{500}=0.51$ Mpc, an increase by a factor of 1.41. For the estimation of $r_{500}$ we use the observed X-ray temperature and scaling relation given by \citet{sun08}. Despite this, as seen in Fig. \ref{sr}, the group is dimmer in X-ray than what is expected from the scaling relations known for the X-ray selected fossils\citep{kpj07}.  

Another group, 2PIGG 2515, has an X-ray luminosity of $L_X=2.7\times 10^{41}{\rm erg\,s}^{-1}$. With the above assumptions on the X-ray surface brightness profile, we obtain an extrapolated bolometric luminosity of $L_X(500)=4.7\times10^{41} {\rm erg\,s}^{-1}$, within $r_{500}=0.34$ Mpc, a factor of 1.75 increase compared to the measured X-ray luminosity.  Within the measurement uncertainty, this group satisfies the fossil X-ray luminosity limit. The same applies to 2PIGG 1635 for which we find an extrapolated X-ray luminosity of $L_X=4.7\times10^{41} {\rm erg\,s}^{-1}$.  However, based on higher quality imaging obtained in this study, this system does not meet the requirement for large luminosity gap and thus can not be classified as a fossil group. 

As outlined in the introduction, the main aim of this study was to test the hypothesis that the collapsed core of the galaxy groups is a suitable environment for the build up of giant elliptical galaxies. Admitting the statistical limitation of the sample, this study supports the hypothesis by demonstrating that groups possessing a giant elliptical BGG, are associated with an extended X-ray emission. A more rigours test requires a statistically larger sample. In this study we demonstrate that the diffuse X-ray emission is detected in 3 out of the 4 targeted groups. For these 3 groups, the observed X-ray luminosity is 2 to 4 times the expected X-ray luminosity from typical galaxy groups. Thus, within the statistical uncertainty, it appears that giant elliptical galaxies form within the collapsed core of galaxy groups.

With these findings and with the caveat that we have only selected galaxy groups with at least 5 spectroscopically confirmed members, we can obtain a rough estimation for the number density of optically identified fossil groups. As described in \citep{jones03} the number density of X-ray selected fossil groups with $L_X\geq 0.5 \times 10^{41} h_{70}^{-2}{\rm erg\,s}^{-1}$ is found to be $\approx 11^{+7.4}_{-4.9}\times 10^{-6}h_{70}^3$ Mpc$^{-3}$, adapted to the cosmology assumed here. This is 10 times the number density of rich galaxy clusters \citep{jones03}. 

The 2PIGG galaxy groups catalog covers a total area of $\approx 1400$ deg$^2$ in north and south stripes of the 2dFGRS and the observed sample of groups in this study is complete out to $z=0.067$ (Table \ref{table1}). The total comoving volume of the survey corresponds to $3\times10^6$ Mpc$^3$, at this redshift. Thus the number density of fossil groups is calculated to be $\approx3-7\times10^{-7}h_{70}^3$ Mpc$^{-3}$, assuming that only 1 (or at most 2) of the observed groups meet the fossil X-ray criteria. This is 20 to 40 times less than the fossil number density found in the WARPS project, which was built on serendipitous X-ray sources from ROSAT Pointed observations \citep{jones03} and 5 to 10 times less than the number density estimate based on the Millennium cosmological simulations \citet{dariush07}, which is reported to be $\approx 35\pm1 \times 10^{-7}h_{70}^3$ Mpc$^{-3}$. However, we note that, as referred to in section 2, the $N_{gal}\geq 5$ criterion only allows a small fraction of  galaxy groups with a large luminosity gap to be selected for X-ray follow-up. Relaxing the restriction on $N_{gal}$, but retaining our other selection criteria described in section 2 (the luminosity gap and the BGG luminosity), we find a total of 48 groups within z=0.067, as compared to 4 groups selected for the X-ray follow-up. Assuming only 25 to 50 percent of the groups meet fossil groups criteria, the number density of fossil groups is estimated to be $4$ to $8 \times10^{-6}h_{70}^3$ Mpc$^{-3}$, nearly equal to the Millennium simulations estimate and only half the space density obtained from the X-ray selected sample \citep{jones03}. We note the statistical limitation of the sample and thus the values should be taken as a rough estimation. 

\begin{figure*}
  \begin{center}
    \leavevmode
      \epsfxsize=10cm\epsfbox{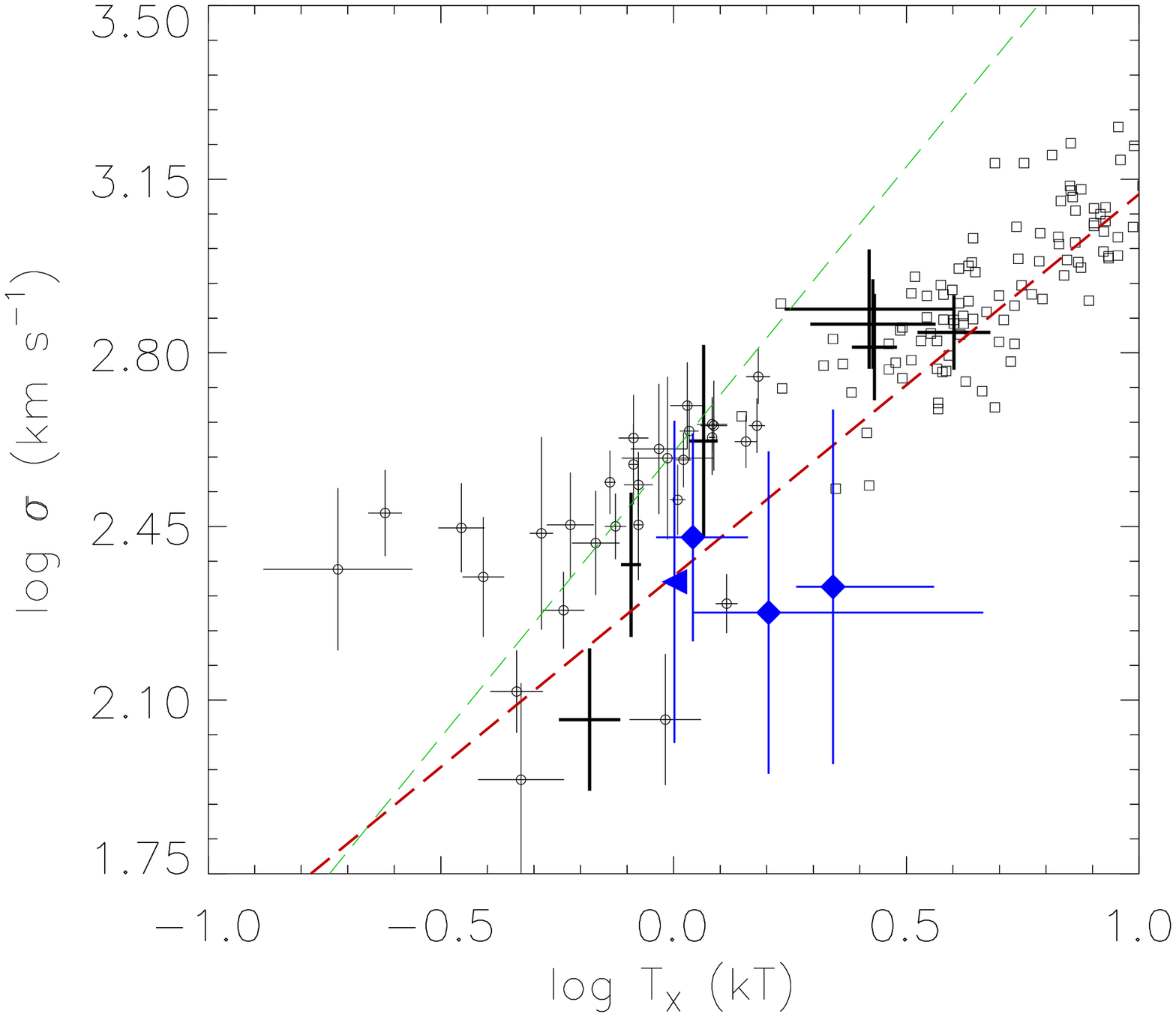}

\caption[composite]{The $T_X:\sigma$ relation for the optically selected groups in this study (dark/blue symbols) and the X-ray selected fossil groups \citep{kpj07} (black bold), GEMS galaxy groups \citep{osmond04} (open circles; best fit green dashed line) and galaxy clusters \citep{wu99}(open squares; best fit red dashed line) . The dark/blue triangle represents 2PIGG 1404 for which we assume 1 keV temperature in spectral fitting. }
     \label{sigmatx}
  \end{center}
\end{figure*}

Finally, to highlight an interesting feature of fossil groups, Fig. \ref{sigmatx} shows a comparison between a sample of galaxy groups \citep{osmond04}, X-ray selected fossils \citep{kpj07} and the observed sample of optically selected groups in the plane of $T_X:\sigma$ relation. As seen, some of our sample groups, the fossils in particular, have a hotter IGM for the group velocity dispersion. This relatively higher IGM temperature observed in optically selected groups could also support the argument that the core of these groups are collapsed. Deeper X-ray observations and complementary spectroscopic membership identification can reveal the true nature of the optically selected groups with a dominant giant elliptical galaxy.

Acknowledgement: We wish to thank Alexis Finoguenov for his useful comments and the anonymous referee for comments and suggestions that helped us improve the presentation and the discussion of this manuscript.

\label{lastpage}


\begin{thebibliography}{99}
 \bibitem[\protect\citeauthoryear{Adami et al.}{2007b}]{b2} Adami,C., Russeil, D., Durret, F., 2007b, A\&A,
  467, 459
  \bibitem[\protect\citeauthoryear{e.g. Alshino et al.}{2010}]{b7} Alshino, A., Khosroshahi,H.,Ponman, T., et al.  2010, MNRAS, 410, 941
\bibitem[\protect\citeauthoryear{Bertin \& Arnouts}{1996}]{b1} Bertin E., Arnouts S. 1996, A\&AS,
 117, 393
\bibitem[\protect\citeauthoryear{Colless} {1989}]{b6} Colless, M. 1989, MNRAS, 237, 799 
\bibitem[\protect\citeauthoryear{Dariush \etal}{2007}]{dariush07}
Dariush A. A., Khosroshahi H. G., Ponman T. J., Pearce F., Raychaudhury S., Hartly W., 2007, MNRAS, 382, 433
\bibitem[\protect\citeauthoryear{Dariush \etal}{2010}]{dariush10}       
Dariush A. A., Raychaudhury S., Ponman T. J., Khosroshahi H. G., Benson A. J., Bower R. G., Pearce F., 2010, MNRAS, 405, 1873
\bibitem[\protect\citeauthoryear{Dunkley et al.}{2009}]{b11} Dunkley J., Komatsu, E., Nolta, M. R., et al.  2009, ApJS, 180, 306
\bibitem[\protect\citeauthoryear{Eke et al.} {2004}]{b5} Eke, V.R.; Baugh C.M.; Cole S., et al. 2004, MNRAS, 348, 866
\bibitem[\protect\citeauthoryear{Ellis \& O'Sullivan} {2006}]{ellis06}  Ellis, S. C.; O'Sullivan, Ewan, et al. 2006, MNRAS, 367, 627
\bibitem[\protect\citeauthoryear{Finoguenov et al.}{2010}] {b8} Finoguenov, A., Watson,M., G., Tanaka, M., et al. 2010, MNRAS, 403, 2063 
\bibitem[\protect\citeauthoryear{Girardi et al.}{1998}]{b12} Girardi M., Giuricin G., Mardirossian F., Mezzetti M., Boschin, W. 1998, ApJ, 505, 74
\bibitem[\protect\citeauthoryear{Gozaliasl et al.}{2013}]{gozali13} Gozaliasl, G, Finoguenov A, Khosroshahi H. G., et al, 2013, submitted to A\&A
\bibitem[\protect\citeauthoryear{Helsdon \& Ponman}{2003}]{helsdon03} Helsdon, S. F., Ponman, T. J., 2003, MNRAS, 340, 485
\bibitem[\protect\citeauthoryear{Kalberla et al.}{2005}]{kalb05}  Kalberla P.~M.~W., Burton W.~B., Hartmann D., Arnal E.~M., Bajaja E., 
  Morras R., P{\"o}ppel W.~G.~L., 2005, A\&A, 440, 775 
\bibitem[\protect\citeauthoryear{Khosroshahi, Jones \&  Ponman}{2004}]{kjp04}  Khosroshahi H.G., Jones L.R., Ponman T.J. 2004, MNRAS, 349, 1240 
\bibitem[\protect\citeauthoryear{Khosroshahi, et al}{2006}]{kmpj06}  Khosroshahi H.G., Maughan B.J., Ponman T.J., Jones L.R. 2006, MNRAS, 369, 1211
\bibitem[\protect\citeauthoryear{Khosroshahi, Ponman \& Jones}{2006}]{kpj06}
Khosroshahi H. G., Ponman T. J., and Jones L. R., 2006, MNRAS Letters, 372, 68
\bibitem[\protect\citeauthoryear{Khosroshahi, Ponman \& Jones}{2007}]{kpj07}  Khosroshahi H.G.,  Ponman T.J., Jones L.R. 2007,  MNRAS, 377, 595
\bibitem[\protect\citeauthoryear{Khochfar \& Burket}{2005}]{khochfar05} Khochfar S., Burkert A., 2005, MNRAS, 359, 1379
\bibitem[\protect\citeauthoryear{e.g. Kodama \&  Arimoto}{1997}]{b3} Kodama,T.,Arimoto, N. 1997, A\&A, 320, 41 
 \bibitem[\protect\citeauthoryear{Jones et al.}{2003}]{jones03} Jones , L. R., Ponman, T. J., Horton, A., et al.  2003, MNRAS, 343, 627
 \bibitem[\protect\citeauthoryear{Lin et al.}{2004}]{b16} Lin, Y. T., Mohr, J. J., Stanford, S. A. 2004, ApJ, 610, 745
 \bibitem[\protect\citeauthoryear{Miller et al.}{2005}]{b15} Miller, C. J., et al 2005, AJ, 130, 968
\bibitem[\protect\citeauthoryear{ Milosavljevic et al.}{2006}]{b14} Milosavljevic, M., Miller C. J., Furlanetto R., Cooray A. 2006, ApJ, 637, 9
\bibitem[\protect\citeauthoryear{Miraghaee et al.}{2013}]{miraghaee13} Miraghaee H., Khosroshahi, H.G., Klockner, H.R., Jetha, N. N., Raychaudhury, S., Ponman, T.J., 2013, MNRAS, submitted 
\bibitem[\protect\citeauthoryear{Osmond \& Ponman}{2004}]{osmond04} 
Osmond J.P. F. and Ponman T. J., 2004, MNRAS, 350, 1511
\bibitem[\protect\citeauthoryear{O'Sullivan, Forbes \& Ponman}{2001}]{osullivan01} 
O'Sullivan E., Forbes D. A., Ponman T. J., 2001, MNRAS, 328, 461
\bibitem[\protect\citeauthoryear{Ponman et al.}{1994}]{ponman94} 
Ponman T. J., Allan D. J., Jones L. R., Merrifield M., MacHardy I. M., 1994, Nature, 369, 462
\bibitem[\protect\citeauthoryear{Popesso et al.}{2005}]{b13} 
Popesso, P., Biviano, A., Bhringer, H., Romaniello, M., Voges, W. 2005, A\&A, 433, 431
\bibitem[\protect\citeauthoryear{Rasmussen et al}{2006}]{jesper06} 
Rasmussen J., Ponman T.J., Mulchaey J. S., Miles T. and Raychaudhury S., 2006, MNRAS, 
\bibitem[\protect\citeauthoryear{Santos \etal}{2007}]{santos07} Santos W. A., Mendes de Oliveira C., Sodre L. Jr., 2007,  AJ, 134, 1551 
\bibitem[\protect\citeauthoryear{Schechter}{1976}]{b10} Schechter P. 1976, ApJ 203, 297
\bibitem[\protect\citeauthoryear{Scharf \etal}{1997}]{Scharf97} 
Scharf C. A., Jones L. R., Ebeling H., Perlman E., Malkan M., Wegner G., 1997, ApJ, 477, 79
\bibitem[\protect\citeauthoryear{Smith et al.}{2010}]{smith10} 
Smith G.P., Khosroshahi H.G., Dariush A.,  Sanderson A. J. R., Ponman T. J. , Stott J. P. , Haines C. P. , Egami E. , Stark D. P. , 2010, MNRAS, 409, 169
\bibitem[\protect\citeauthoryear{Springel et al.}{2005}]{springel05} 
Springel et al., 2005, Nature, 435, 629
\bibitem[\protect\citeauthoryear{Sun et al.}{2004}]{sun04} 	
Sun, M.; Forman, W.; Vikhlinin, A.; Hornstrup, A.; Jones, C.; Murray, S. S., 2004, ApJ 612 ,  805
\bibitem[\protect\citeauthoryear{Sun et al.}{2004}]{sun08} 	
Sun, M.; Voit, G. M., Donahue M, Jones C., Forman W., Vikhlinin A., 2008, ApJ 693 ,  1142
\bibitem[\protect\citeauthoryear{Tavassoli et al.}{2010}]{b17} 
Tavasoli, Saeed; Khosroshahi, Habib G.; Koohpaee, Ali; Rahmani, Hadi; Ghanbari, Jamshid, 2011,PASP,123,1
 \bibitem[\protect\citeauthoryear{Turner \& Gott}{1976}]{b18} Turner, E.L. \& Gott, J.R., 1976, ApJ, 209, 6
\bibitem[\protect\citeauthoryear{Ulmer et al.}{2005}]{ulmer05} 
Ulmer, M. P.; Adami, C.; Covone, G.; Durret, F.; Lima Neto, G. B.; Sabirli, K.; Holden, B.; Kron, R. G.; Romer, A. K., 2005, ApJ 624 ,  124
\bibitem[\protect\citeauthoryear{Wechsler \etal}{2002}]{wechsler02} 
Wechsler, Risa H.; Bullock, James S.; Primack, Joel R.; Kravtsov, Andrey V.; Dekel, Avishai
\bibitem[\protect\citeauthoryear{Wu, Xue \& Fang}{1999}]{wu99} 
Wu X.-P., Xue H., \& Fang L.-Z. ,1999, ApJ, 524, 22
 \end{thebibliography}
\end{document}